\documentstyle{l-aa}

\def\Ref{\bibitem{}}
\def\eq#1{\begin{equation} #1 \end{equation}}
\def\eqarray#1{\begin{eqnarray} #1 \end{eqnarray}}
\def\non{\nonumber \\ }
\def\D#1:#2/#3  {\if!#1!{d#2 \over d#3} \else {d^{#1}#2 \over d#3^{#1}} \fi}
\def\about{\ifmmode\sim\else$\sim$\fi}
\def\symbol#1{\hbox{$#1$}}
\def\Frac#1/#2{\symbol{\textstyle {#1 \over #2}}}
\def\half   {\Frac 1/2}
\def\third  {\Frac 1/3}
\def\fourth {\Frac 1/4}
\def\x      {\symbol{\times}}

\def\Ri      {\symbol{R_i}}
\def\Rs      {\symbol{R_s}}
\def\xi      {\symbol{x_i}}
\def\xii     {\symbol{x_i^2}}
\def\xs      {\symbol{x_s}}
\def\xss     {\symbol{x_s^2}}
\def\rs      {\symbol{r_s}}
\def\teto    {\symbol{\theta_0}}
\def\dtet    {\symbol{\delta\theta}}
\def\tetb    {\symbol{\theta_b}}
\def\tetbb   {\symbol{\theta_b^2}}
\def\tets    {\symbol{\theta_s}}

\def\lo      {\symbol{\ell_0}}
\def\lI      {\symbol{\ell_1}}
\def\lII     {\symbol{\ell_2}}
\def\k       {\symbol{\kappa}}
\def\ko      {\symbol{\kappa_0}}
\def\Io      {\symbol{I_0}}
\def\Js      {\symbol{J_s}}
\def\Ao      {\symbol{A_{\rm obs}}}

\begin{document}

\thesaurus{02         
           (02.13.3;  
            02.18.7), 
            08.03.4}  

\title
                              {Shell masers}

\author{Moshe Elitzur \inst{1} \and Valent\'{\i}n Bujarrabal \inst{2} }

\offprints{V. Bujarrabal}

\institute{Department of Physics and Astronomy, University of Kentucky,
           Lexington, KY 40506, USA; moshe@pa.uky.edu
           \and
           Centro Astron\'omico de Yebes (IGN), Apartado 148,
           19080 Guadalajara, Spain; bujarrabal@cay.es }

\date{Received April 25, 1995; accepted August 3, 1995}

\maketitle
\markboth{Elitzur \& Bujarrabal: Shell masers}{}

\begin{abstract}

We present the analytical solution of a maser shaped like a spherical shell.
We determine the general condition on the size of the central cavity at which
a sphere becomes a shell maser, and derive the intensity, beaming angle and
observed size of both unsaturated and saturated shells.

\end{abstract}

\section
                                {Introduction}

Given a pumping scheme, the brightness temperature of an unsaturated maser can
be immediately determined from the source's length along the line of sight
because the gain, $\tau$, is simply proportional to it; there is no need to
even consider the geometrical shape.  In a saturated maser, on the other hand,
the dependence of gain on distance traveled through the source varies between
the saturated and unsaturated domains.  As a result, the brightness
temperature in any given direction cannot be determined before a complete
solution of the full maser structure has been attained.  An analytic solution
of this highly nonlinear problem is possible because, in all sources with
an appreciable gain, maser radiation is highly beamed (e.g.\ Elitzur 1990b;
hereafter E90): at any location in an arbitrarily shaped maser, the longest
chord through the source provides the direction of the local dominant ray and
maser radiation is mostly confined to a small beam around it. The complete,
detailed solution is highly dependent on the source geometry; indeed, it
cannot even be attempted without fully specifying the geometry. Thus far,
detailed solutions have been derived for linear masers (Elitzur 1990a), for
the planar geometry of a disk (Elitzur et al.\ 1992), and for the three
dimensional configurations of spheres and filaments (Goldreich \& Keeley 1972,
E90, Elitzur et al.\ 1991).

 \begin{figure}[htbp]
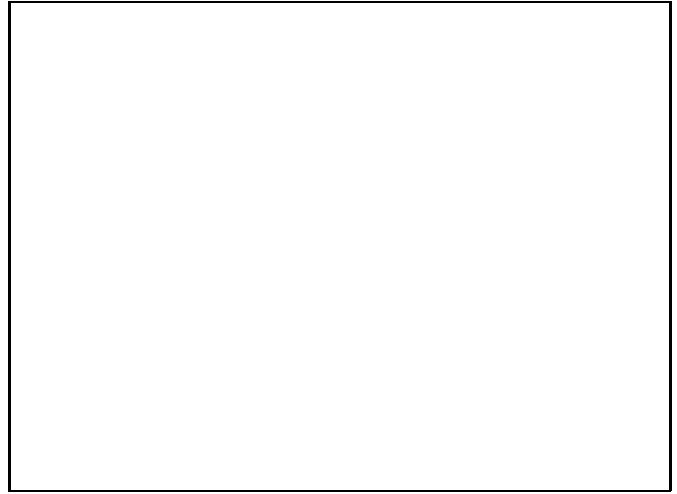

 \centering \leavevmode
 \picplace{6.5cm}
 \caption{Geometrical notations for a shell maser.}
 \end{figure}

Recent interferometric observations of masers in late-type stars reveal
ring-like structures, indicative of a shell geometry.  The observations
include both H$_2$O (Reid \& Menten 1990) and SiO (Diamond et al.\ 1994,
Miyoshi et al.\ 1994, Greenhill et al.\ 1995). The shell geometry is
fundamentally different from all the maser geometries considered until now,
which will be referred to as ``regular''.  In regular geometries there is a
single dominant ray at virtually every point (the only exception is an
occasional singularity, such as the center of a sphere or a disk maser). The
dominant ray is aligned with the local axis of symmetry and its direction
coincides with that of the flux vector.  In contrast, at an arbitrary point in
a shell there are many chords of maximal length belonging to the rays that
graze the core (see Fig.~1). Thus the dominant rays in a shell maser span the
surface of a cone whose axis is aligned with the radius vector. From symmetry,
the flux vector points in the direction of that axis and it does not coincide
with the direction of any dominant ray.

Maser emission from shells has been studied thus far only in numerical
calculations (Western \& Watson 1983; Bujarrabal 1994).  Utilizing the
techniques developed in the studies of regular geometries, we present here the
complete analytical solution of shell masers.  Since the solution methods
are described in detail in the referenced literature, the presentation here
centers mostly on the features unique to the shell geometry.

\section
                                {Generalities}

Consider a maser shaped as a spherical shell with inner radius \Ri\ and outer
radius $R$ (Fig.~1).  As in all previous solutions, the shell is assumed
uniform and quiescent (no ordered motions). At a point in the shell
characterized by radius $r$, the longest chords belong to the rays that graze
the core, inclined at $\teto = \sin^{-1}(\Ri/r)$ to the radius vector.
Introduce dimensionless variables $x$ and \xi\ such that
 \eq{
      r = xR\, , \qquad  \Ri = \xi R\, ,  \qquad \teto = \sin^{-1}(\xi/x).
 }
That is, \xi\ is the relative size of the cavity, and positions in the shell
are characterized by $\xi \le x \le 1$.

Denote the intensity along a dominant ray of any maser \Io.  In regular
geometries the radiation is confined mostly to $|\theta| \le \tetb$, where
$\theta$ is measured from the local symmetry axis and \tetb\ $(\ll 1)$ is the
beaming angle. Therefore, the angle-averaged intensity is $J \simeq
\fourth\Io\tetbb$, if the intensity is assumed approximately constant inside
the maser beam.  In shell masers, on the other hand, the radiation is confined
mostly to $\teto \le \theta \le \teto + \tetb$ (see Sec.~3 below).  Neglecting
rays outside the main beam, the angle-averaged intensity for a constant
intensity inside the beam is
 \eq{
     J = \half\int I\sin\theta d\theta
  \simeq \half\Io\tetb\left(\sin\teto + \half\tetb\cos\teto\right).
}
Instead of a rectangular intensity profile we can employ an exponential
fall-off, $I \propto \exp[-(\theta - \teto)/\tetb)]$.  This angular
distribution produces the same result except that the factor \half\ inside
the parentheses is replaced by unity.

When $\teto \ll \half\tetb\ (\ll 1)$, the second term in the parentheses
dominates and the result of regular geometries is recovered. Therefore, the
shell is properly described by the spherical maser solution as long as its
central cavity is sufficiently small that $\teto < \half\tetb$. The shell
structure must be considered only in the reverse situation in which the first
term in the parentheses dominates and the spherical solution no longer
applies.  So the transition to a shell maser occurs when
 \eq{\label{shell0}
                             \teto > \tetb\, ,
}
omitting the factor \half\ to accommodate also the exponential profile. This
condition, similar to the filamentary condition that defines elongated masers
(Elitzur et al.\ 1991), will be referred to as the {\em shell condition}.  At
the inner surface ($x = \xi$), $\teto = \pi/2$ and the condition is trivially
obeyed.  It provides the most meaningful constraint at the outer surface ($x =
1$), where \teto\ is minimal.

The meaning of the shell condition is quite simple.  On the surface of a
spherical maser, the radiation fills a cone with an opening angle \tetb\
centered on the radius vector. Now begin to carve a hole at the center of the
sphere.  As long as that hole is contained in the radiation cone it can be
ignored and the maser remains a sphere.  The maser becomes a shell once the
hole emerges from the side of the radiation cone, i.e., when the shell
condition is obeyed.  The angle-averaged intensity in a shell maser is thus
 \eq{\label{J}
        J = \half\Io\tetb\sin\teto = \half\Io\tetb{\xi\over x}\, .
}
To a distant observer, the maser appears as a ring-like structure with inner
radius \Ri\ and outer radius $\Ri + H$, where
 \eq{\label{H}
                        H = R\tetb\sqrt{1 - \xii}
}
and \tetb\ is the beaming angle at the surface.  Observations can directly
determine only the inner radius \Ri\ and the thickness $H$.  The outer radius
$R$ is unobservable, thus the ratio \xi\ cannot be directly determined.  This
problem exists of course in all regular geometries, too, where the only maser
dimension amenable to observations is the observed size in the plane of the
sky, not the length along the line of sight.  Neither the radius of a
spherical maser nor the length of a filamentary maser nor the length along the
line of sight of a planar maser are directly observable (e.g.\ Elitzur et al.\
1992).

Consider now the maser flux.  In regular geometries the flux vector points in
the direction of the dominant ray and is dominated by the contribution of
$\theta = 0$, thus its magnitude obeys $F \simeq 4\pi J$.  In a shell maser,
on the other hand, the directions of the dominant ray and flux vector are
different. Since the flux is dominated by the contribution of $\theta =
\teto$, its magnitude obeys
 \eqarray{\label{flux}
    \lefteqn{F = \int\cos\theta Id\Omega \simeq \cos\teto\int Id\Omega} \non
           & & = \cos\teto\x4\pi J = 4\pi J \sqrt{1 - (\xi/x)^2}.
 }
The result for regular geometries is recovered when \xi\ = 0.  The flux
vanishes on the shell's inner surface because non-dominant rays were
neglected.

\section
                          {Unsaturated shell masers}

The optical depth in an unsaturated maser is always proportional to pathlength.
At an arbitrary position in the shell, denote the length of the longest chord
\lo.  A ray inclined to this chord at a small angle \dtet\ ($> 0$) has a
length
 \eqarray{
 \lefteqn{\ell = r\cos(\teto + \dtet) + \sqrt{R^2 - r^2\sin^2(\teto +
\dtet)}}\non
           & & \simeq \lo - \lI\dtet - \lII\dtet^2,
 }
where
 \eqarray{
   \lo  &=&      R\left(\sqrt{x^2 - \xii} + \sqrt{1 - \xii}\right)        \non
   \lI  &=&    \Ri\left(1 + \sqrt{{x^2 - \xii \over 1 - \xii}}\right)     \\
   \lII &=&\half R\left[\sqrt{x^2 - \xii} +
                 {x^2 - 2\xii + x_i^4 \over (1 - \xii)^{3/2}}\right]. \nonumber
 }
In regular geometries the dominant ray corresponds to a true maximum of the
geometrical surface, i.e., $d\ell/d\theta = 0$ and the variation of $\ell$
with \dtet\ is quadratic.  In a shell the dominant ray does not correspond to
a true maximum, the first derivative does not vanish and the variation of
$\ell$ with \dtet\ includes a linear term (which properly vanishes when \Ri\ =
0). As a result, the angular distribution in an unsaturated shell is
 \eq{\label{angular}
               I(\dtet) = I_0 e^{-\ko\lI\dtet - \ko\lII\dtet^2}.
 }
Here \ko\ is the unsaturated absorption coefficient\footnote{Following what
is by now a standard procedure, the negative of the absorption coefficient
is used for the maser transition so that it is positive.} and $I_0 =
S\exp(\ko\lo)$, where $S$ is the source function.  The intensity detected by
an external observer is obtained by substituting $x = 1$, which produces $\lo
= 2\lII = 2R\sqrt{1 - \xii}$ and $\lI = 2\Ri$.  Instead of the standard
Gaussian that typifies the angular distribution of spherical maser radiation,
the argument of the exponential contains a term linear in \dtet.  At very
small \dtet\ the linear term dominates, at very large \dtet\ the quadratic
term.  The shell condition ensures that the linear term dominates inside the
entire maser beam, $\dtet \le \tetb$.  With the dominance of this term, the
radiation at any point in an unsaturated shell maser varies with direction in
proportion to exp($-\dtet/\tetb$), where
 \eq{\label{tetb-uns}
                        \tetb = {1 \over \ko\lI}
 }
is the beaming angle. At the shell's outer surface $\tetb = 1/(2\ko\Ri)$, and
the shell condition becomes
 \eq{\label{shell}
                          \xii > {1\over\ko R}\, .
 }
A factor of \half\ was omitted on the right-hand side to provide a slightly
more stringent constraint, in line with another estimate described below.

Only rays with $\dtet > 0$ were considered thus far.  Rays with $\dtet < 0$
cut through the cavity and the decrease in their length from \lo\ is
dominated by a term proportional to $\sqrt{-\dtet}$, much steeper than the
linear decline for $\dtet > 0$.  For all practical purposes, the angular
distribution can always be assumed to have a sharp cutoff so that only
$\theta \ge \teto$ need be considered.

The condition just derived was determined for unsaturated shells.  However, a
maser cannot become saturated unless it was first unsaturated and had to obey
this condition during its unsaturated phase to be considered a shell.
Therefore, Eq.~\ref{shell} is the general condition that must be obeyed by all
shell masers.  For saturated shells we have to consider the magnitude of $\ko
R$ that the maser must have had to become saturated. Typically, saturation
occurs at $\ko R \ga 10$, thus astronomical masers obey the shell condition
when $\xi \ga 0.3$.

General results for arbitrary \xi, displaying explicitly the transition from
a sphere to a shell, can be obtained from the full angular distribution of
Eq.~\ref{angular}. The requirement that the intensity drop to $1/e$ of its
peak value produces a quadratic equation for the beaming angle whose solution
is
 \eq{\label{tetb full}
            \tetb = {\lI \over 2\lII}\left(\sqrt{1 + z^{-2}} - 1 \right),
 }
where $z = \ko\lI/(2\sqrt{\ko\lII})$. When $z \gg 1$, equivalent to the shell
condition of Eq.~\ref{shell}, the beaming angle of Eq.~\ref{tetb-uns} is
recovered.  The opposite limit, $z \to 0$, properly reproduces the beaming
angle of an unsaturated sphere, $\tetb = 1/\sqrt{\ko R}$.  Furthermore, the
complete expression for $J$ can be calculated from an integration of the full
angular distribution, and the result is
 \eq{
 J = \fourth\Io{1\over\ko\lII}\left[1 - \sqrt{\pi}(z - \teto\sqrt{\ko\lII})
                                               e^{z^2} \hbox{erfc}\ z \right].
 }
Again, utilizing properties of the error function erfc at large and small $z$
reproduces the proper expressions for $J$ in the respective limits.

{}From Eq.~\ref{H}, the observed thickness of the maser ring obeys
 \eq{
                   \ko H = {\sqrt{1 - \xii} \over 2\xi}\, .
 }
This result can also be obtained by noting that the surface brightness
observed at infinity varies as $e^{-h/H}$, where $h$ is the parallel linear
displacement from the brightest ray.  It is interesting that the observed
thickness of unsaturated shells is independent of size,\footnote{The maser
size enters only indirectly through the condition $\ko R \gg 1$, required to
ensure that the maser provides amplification sufficient to cause beaming.}
the shell dimensions enter only in the ratio $\Ri/R$.

Since the thickness $H$ was derived assuming the shell condition, it diverges
when $\xi \to 0$.  Employing the full expression for the beaming angle,
Eq.~\ref{tetb full}, removes the divergence and properly produces $H =
\sqrt{R/\ko}$ when \xi\ = 0, the observed size of an unsaturated spherical
maser.  In addition, the area generally provides a useful indicator for the
effects of geometry on dimensions of the observed maser (see table 1 in
Elitzur et al.\ 1992).  The observed area of an unsaturated shell maser, where
$\Ao = 2\pi\Ri H$,  is
 \eq{
                       A_{\rm obs} = {\pi\lo\over2\ko}\,,
 }
where \lo\ is the longest chord at the outer surface. By comparison, in an
unsaturated sphere, where $\lo = 2R$ and the observed region is a small cap,
$\Ao = 2\pi\lo /\ko$. Apart from a factor of 4, the expressions for the
observed areas are the same.

\section
                           {Saturated shell masers}

Consider now a sequence of masers with a fixed \xi, increasing $\ko R$.  If the
original maser obeys the shell condition, Eq.~\ref{shell}, subsequent masers
obey it with a larger margin.  The emerging intensity grows exponentially with
maser size and at a certain radius, \Rs, it obeys $J(x = 1) = \Js$, i.e., the
shell saturates.  The corresponding radius obeys
\eq{
     {1\over\ko\Rs}\exp\left(2\ko\Rs\sqrt{1 - \xii}\right) = 4\gamma,
}
where $\gamma = \Js/S$; typically, astronomical masers have $\gamma \about
10^5 - 10^7$.  The approximate solution of this equation is
 \eq{
 \ko\Rs \simeq
  {1\over 2\sqrt{1 - \xii}}\ln\left({2\gamma\over\sqrt{1 - \xii}}\right),
 }
a solution valid when the argument of the logarithm on the right-hand-side is
large (e.g.\ E90). When \xi\ = 0, this result reproduces the saturation radius
of a spherical maser.

With further increase in radius, the shell develops a two-zone structure: the
inner zone, $\Ri \le r \le \rs$, is unsaturated and has $\k \simeq \ko$, the
outer zone, $\rs \le r \le R$, is saturated and has $\k \simeq \ko\Js/J$. The
boundary radius, $\rs \equiv \xs R$, is still unspecified.  It is \Rs\ for a
maser that has just saturated and should shrink logarithmically as $R$
increases.

\subsection                 {The saturated shell}

{}From Eq.~\ref{flux}, the flux divergence relation in a shell maser is
\eq{\label{eqJ}
           \D:{}/x \left(x\sqrt{x^2 - \xii}J\right) = \kappa JRx^2.
 }
Since in the saturated shell $\kappa J \simeq \ko\Js$, the solution there is
immediate:
\eq{
 x\sqrt{x^2 - \xii}J - \xs\sqrt{\xss - \xii}\Js = \third\Js\ko R(x^3 - x_s^3).
 }
Note that when $x \gg \xs$, $J = \third\Js\ko r$, same as the result for a
sphere (E90).

Along a ray that grazes the core, $I = \Io$ and $d\ell = d\lo$.  Thus the
radiative transfer equation past the mid-point of this path is
 \eq{
             \sqrt{1 - {\xii\over x^2}} \D:{\Io}/x = \kappa R\Io.
 }
With the aid of Eq.~\ref{J}, the transfer equations for $J$ and \Io\ produce an
equation for the beaming angle
 \eq{
       \D:{\ln(\tetb\xi/x)}/x = - {2x^2 - \xii \over x(x^2 - \xii)}\,.
 }
The solution is straightforward, the result is
 \eq{
             \tetb = \tets\sqrt{\xss - \xii \over x^2 - \xii}\,,
 }
where \tets\ is the beaming angle at $x = \xs$; again, \xi\ = 0 reproduces the
expression valid in the saturated zone of a spherical maser (E90).  The angle
\tets\ can be obtained from the result for an unsaturated maser
(Eq.~\ref{tetb-uns}), yielding
 \eq{
               \tets = {1\over\ko R\xi}{\sqrt{1 - \xii} \over
      \sqrt{1 - \xii} + \sqrt{\xss - \xii}} \simeq {1\over\ko\Ri}\,.
 }
In the last relation we assumed that the thickness of the saturated zone
greatly exceeds that of the unsaturated region, i.e., $1 - x_i \gg x_s - x_i$,
an assumption justified for strongly saturated shells.  The thickness of the
observed maser ring is
 \eq{
                         H = \tets\sqrt{r_s^2 - R_i^2}
}
and its area is
 \eq{
                  \Ao = {2\pi\over\ko}\sqrt{r_s^2 - R_i^2}\,.
}
By comparison, the observed area of a saturated sphere is $\pi r_s/\ko$.

The final quantity needed to complete the solution is \rs, the radius of the
boundary between the saturated and unsaturated zones.  It can be obtained
from the requirement that, amplification by the unsaturated shell brings the
intensity of the subordinate ray at core entrance to the dominant ray
corresponding to \Js\ upon exit. The detailed steps are described in E90 and
will be omitted here.  The final result is controlled by the parameter
 \eq{\label{b}
      b = \ln\left({12\gamma\xii \over (\ko R)^2\sqrt{1 - \xii}}\right).
}
When the argument of the logarithm is much larger than unity, which is usually
the case in astronomical masers, the boundary of the unsaturated zone obeys
 \eq{\label{rs}
             (\ko\rs)^2 \simeq (\ko R_i)^2 + \fourth b^2 .
}
This enables us to complete the solution for a saturated shell maser.
The beaming angle of the emergent radiation is
 \eq{
              \tetb = {b\over 2(\ko R)^2 \xi\sqrt{1 - \xii}}\,,
}
its intensity is
 \eq{
                 I_0 = {4\over3b}\Js(\ko R)^3\sqrt{1 - \xii}
}
and the thickness of the observed maser ring obeys
 \eq{
                         \ko H = {b \over 2\ko\Ri}\,.
}

\subsection                 {Complete saturation}

As $\ko R$ increases while \xi\ is held fixed, the parameter $b$ decreases
logarithmically and $r_s \to R_i$ (see Eqs.~\ref{b} and \ref{rs}); that is, the
size of the unsaturated zone shrinks.  In analogy with core saturation of
regular geometries, the unsaturated zone disappears when $R$ exceeds a certain
threshold $R_c$ and the shell is saturated throughout.  A reasonably accurate
estimate for the radius $R_c$ can be obtained by setting the argument of the
logarithm in Eq.~\ref{b} to unity, yielding
 \eq{
             \ko R_c \simeq 3.5\gamma^{1/2}\xi(1 - \xii)^{-1/4}\,.
 }
Complete saturation requires dimensions that greatly exceed typical sizes of
astronomical masers, thus it is mostly only of theoretical interest.

\section
                                 {Discussion}

A shell becomes a sphere when $\xi \to 0$. Some of the expressions we list are
based on the shell condition and diverge in that limit.  These divergences
can be avoided by employing the full expression for the unsaturated beaming
angle (Eq.~\ref{tetb full}) instead of its shell limit (Eq.~\ref{tetb-uns}).
The results display explicitly the transition between the two geometries at
the price of greater complexity.

While a frequency dependence was never indicated explicitly, it enters
through the frequency variation of \ko, the unsaturated absorption
coefficient. The full absorption coefficient, including saturation effects,
used in this study was assumed to obey the standard form proposed by Goldreich
\& Kwan (1974).  This implies that our results are valid around line center up
to frequency shifts of at least two Doppler widths (Elitzur 1994), covering
practically all cases of interest.  A more significant limitation is that, as
with all maser solutions available to date, the one presented here is valid
only for quiescent material.  Complete analytic solutions for masers with
large velocity gradients do not yet exist for any geometry. While the impetus
for this study comes from the SiO interferometry results which indicate
ring-like structures, these observations also show that the maser emission
comes from many compact spots, each one characterized by a different Doppler
velocity (Diamond et al.\ 1994, Miyoshi et al.\ 1994, Greenhill et al.\ 1995).
 Each feature may thus correspond to a clump, defined either by a local
enhancement of the pump rate (reflecting a favorable combination of conditions
such as density, abundance, temperature, etc.) or velocity coherence in a
turbulent medium.  In either case, the geometry relevant to the maser
radiation is the local one of the individual features, perhaps best described
by elongated structures. On the other hand, the significant amount of flux
lost in VLBI measurements due to overresolution indicates that about one half
of the total maser emission is much more extended than the VLBI spots. It is
also known from measurements with lower spatial resolution that the total
emission spreads over a region comparable in size to the ring seen in VLBI
data. Indeed, if the distribution of a component of the emission smoothly
fills up such a ring, it would very probably not be detectable in the high
resolution experiments. Such extended masers would be described by the
solution presented here.

The absence of observational data on such possible extended emission prevents
further discussion at this time. As is the case for the spherical geometry
with which it is closely associated, the primary importance of the shell maser
solution may be in the insight it offers to maser behavior rather than direct
comparison with current observations.

\begin{acknowledgements}
We thank the referee, Dr.\ W.H.\ Kegel, for his careful reading of the
manuscript and useful comments and suggestions, especially for pointing out
the general expression for the beaming angle, Eq.~\ref{tetb full}. This work
was begun during a visit to Centro Astronomico de Yebes by M.E. He would like
to thank the Center for its hospitality and the NSF for its partial support
through grant AST-9321847. V.B.\ acknowledges partial support by DGICYT
through grant PB93-0048.

\end{acknowledgements}

\end{document}